\documentclass[twocolumn,superscriptaddress,nofootinbib,floatfix,aps,showpacs,prb,citeautoscript,reprint]{revtex4-1}
\usepackage{dcolumn}
\usepackage{bm}
\usepackage{amsmath}
\usepackage{amsfonts}
\usepackage{amssymb}
\usepackage{graphicx}
\usepackage{tabularx}
\usepackage{subfigure}
\usepackage{color}
\usepackage{gensymb}
\usepackage{physics}
\usepackage{hyperref} \hypersetup{colorlinks=true, linkcolor=blue,
  citecolor=blue, urlcolor=blue, pdftitle={Prediction of Weyl semi-metallic phase in inversion-asymmetric BiSb}}

\begin{document}
\title{Prediction of a controllable Weyl semi-metallic phase in inversion-asymmetric BiSb}

\author{Sobhit Singh}
\affiliation{Department of Physics and Astronomy, West Virginia University, Morgantown, WV-26505-6315, USA}

\author{A. C. Garcia-Castro}
\affiliation{Centro de Investigaci\'on y Estudios Avanzados del IPN, MX-76230, Quer\'etaro, M\'exico}
\affiliation{Physique Th\'eorique des Mat\'eriaux, Universit\'e de Li\`ege, B-4000 Sart-Tilman, Belgium}

\author{Irais Valencia-Jaime}
\affiliation{Department of Physics and Astronomy, West Virginia University, Morgantown, WV-26505-6315, USA}
\affiliation{Centro de Investigaci\'on y Estudios Avanzados del IPN, MX-76230, Quer\'etaro, M\'exico}

\author{Francisco Mu\~noz}
\affiliation{Departamento de F\'isica, Facultad de Ciencias, Universidad de Chile, Casilla 653, Santiago, Chile}
\affiliation{Centro para el Desarrollo de la Nanosciencia y la Nanotechnolog\'ia CEDENNA, Santiago, Chile}

\author{Aldo H. Romero}
\affiliation{Department of Physics and Astronomy, West Virginia University, Morgantown, WV-26505-6315, USA}

\begin{abstract}
Recent experimental realization of long sought Weyl fermions in non-magnetic crystals has greatly motivated condensed matter physicists to search for materials supporting Weyl fermions. Weyl fermions appear to be very promising for future electronics, often referred as Weyltronics. Here, by means of first-principle calculations, we report a stoichiometric crystal structure of BiSb with broken space-inversion symmetry. This structure is insulating in bulk and has non-trivial band topology. We observe a pressure driven Weyl semi-metallic phase transition in this crystal structure. The obtained Weyl semi-metallic phase exists in the 4.0 - 6.0 GPa pressure range. We find that a total 6 pairs of Weyl points, 6 monopoles and 6 antimonopoles, exist in the Brillouin zone. The Weyl points with opposite chirality are located at different energy values yielding separate electron and hole Fermi-surfaces. Additionally, the spin-texture of the bulk BiSb compound appears to be electrically controllable when the interlink between pressure and an electric field is exploited. This produces novel manipulable topological transport properties in this system which are very promising for implementation of this kind of materials in next-generation Weyltronics and spintronic devices.
\end{abstract}


\maketitle

Weyl fermions have recently attracted the attention of researchers due to their unique intriguing physical properties such as the presence of discontinuous Fermi-arcs \cite{WanPRB2011, WeylKiss2014, Potter2014, XuScience2015}, quantized anomalous Hall effect \cite{GXuPRL2011, Zyuzin2012}, quantum transport \cite{WangPRB2013}, anomalous magnetoresistance \cite{KimPRL2013}, etc. Weyl fermions are an alternative massless solution of the two-component Dirac equation, which is also known as the Weyl equation \cite{Weyl1929}. These fermions were initially hypothesized by particle physicists to explain the chiral and massless behavior of neutrinos (assuming neutrinos have negligible mass). In condensed matter systems, Weyl fermions can be realized as low-energy excitations near the touching points of two non-degenerate bands at discrete $\textbf{k}$-points in the momentum space of a bulk system. These gapless band touching points are known as Weyl nodes or Weyl points \cite{WeylKiss2014}. Close to a Weyl point, bands have linear dispersion in all directions in $\textbf{k}$-space. In 3D, Weyl points are similar to the well-known Dirac points except for the fact that Weyl points are spin non-degenerate, but Dirac points have spin-degeneracy. By breaking either the time-reversal (TR) symmetry or space-inversion (I) symmetry, one can lift up the spin-degeneracy of the Dirac points in a Dirac semimetal \cite{YoungPRL2012} and realize a Weyl semimetal (WSM). Hence, the lack of TR-symmetry or I-symmetry is an essential requirement for realization of WSM phase in 3D systems. 

Each Weyl point can be characterized by an associated chirality (left handedness or right handedness) or by a monopole and an antimonopole having charge $+1$ or $-1$, respectively. Weyl points are always created in pairs via pair-creation of a monopole and an antimonopole \cite{ZyuzinPRB2012, WeiPRL2012, Vanderbilt2014}. Similarly, they can disappear only via the pair-annihilation process. Thus, the total charge in $\textbf{k}$-space is always zero and the total number of Weyl points is always even. In I-symmetric WSMs, opposite Weyl points in a Weyl-pair are located at the same energy while in I-asymmetric WSMs opposite Weyl points are separated both in the momentum space and in energy. This eliminates the possibility of realization of a nodal semimetallic state by tuning the Fermi-level and thus, separate electron and hole Fermi-surfaces exist in the I-asymmetric WSM systems. Such systems are very interesting and possess novel topological transport properties \cite{Fukushima2008, Grushin2012, Aji2012, ZyuzinPRB2012, Zyuzin2012, Hosur2013, Vazifeh2013, Chang2015}.

In recent years, many different theoretical proposals have been reported for realization of WSM phase in I-symmetric \cite{WanPRB2011, Yang2011, BalentsPRL2011, BalentsPRB2011} and I-asymmetric systems \cite{Murakami2007, BalentsPRB2012, Wang2012, SinghPRB2012, Ojanen2013, Vanderbilt2014, HirayamPRL2015, WengPRX2015, soluyanov2015type, Huang02022016}. Several experimental groups have recently reported the experimental realization of WSM phase in single crystal compounds \cite{Huang2015, xu2015discovery, Lv2015, XuS2015} and in photonic crystals \cite{gyroid2015}. To the best of our knowledge, there is no report of controllable WSM phase in a compound. However, ferroelectric-like polar WSM compounds having large spin-orbit coupling (SOC) could open a new avenue of realization of novel transport properties. When we combine SOC with ferroelectricity, we obtain a class of novel multifunctional materials, called Ferro-Electric Rashba Semiconductors (FERSC) \cite{Silvia2014}. Due to the giant Rashba spin-splitting present in FERSCs, the electric-field control of the spin-texture is possible in FERSCs. Recently, S. Picozzi \emph{et al.} have theoretically demonstrated the possible control and switching of the spin-texture by means of an electric field in GeTe, which is a ferroelectric compound inheriting large SOC \cite{Sante_GeTe2013, Silvia2014}. This prediction was later experimentally corroborated \cite{Marcus2016}. In addition to GeTe compound, a possible control of the spin-texture and spin-valley-tronics via an electric-field has been predicted for tin iodine perovskites (\emph{FA})SnI$_3$ \cite{Stroppa2014}  ---with \emph{FA} = (NH$_2$CHNH$_2$)$^+$--- and BiAlO$_3$/BiIrO$_3$ (111)/(111) superlattices \cite{PhysRevLett.115.037602}, respectively.

The Bi$_{1-x}$Sb$_x$  alloys are the first generation topological insulator \cite{KaneRev2010}. In 2013, H. J. Kim \emph{et al.} reported experimental evidence of WSM phase in Bi$_{1-x}$Sb$_x$ alloys for $x$ = 3$\%$ by magnetoresistance measurements \cite{KimPRL2013}. In the present work, we report a stoichiometric crystal structure of BiSb with broken I-symmetry. This structure belongs to the space group 160 (\emph{R3m}) and has non-trivial band topology. By means of first-principle calculations, we observe a pressure-induced WSM phase transition in the present BiSb system in 4.0 - 6.0 GPa pressure range. We have further exploited the coupling between large SOC and ferroelectricity of BiSb to tune the spin-texture of Rashba-like bands by an electric field. Our first-principle calculations show that the chirality of the Weyl points in BiSb can be switched by an electric field. This property of a ferroelectric-like polar WSM is very important and such electric-field control of Weyl points could play a vital role in the next generation Weyltronics and spintronics devices. 

\begin{figure}[htb!]
 \centering
 \includegraphics[width=8.8cm, keepaspectratio=true]{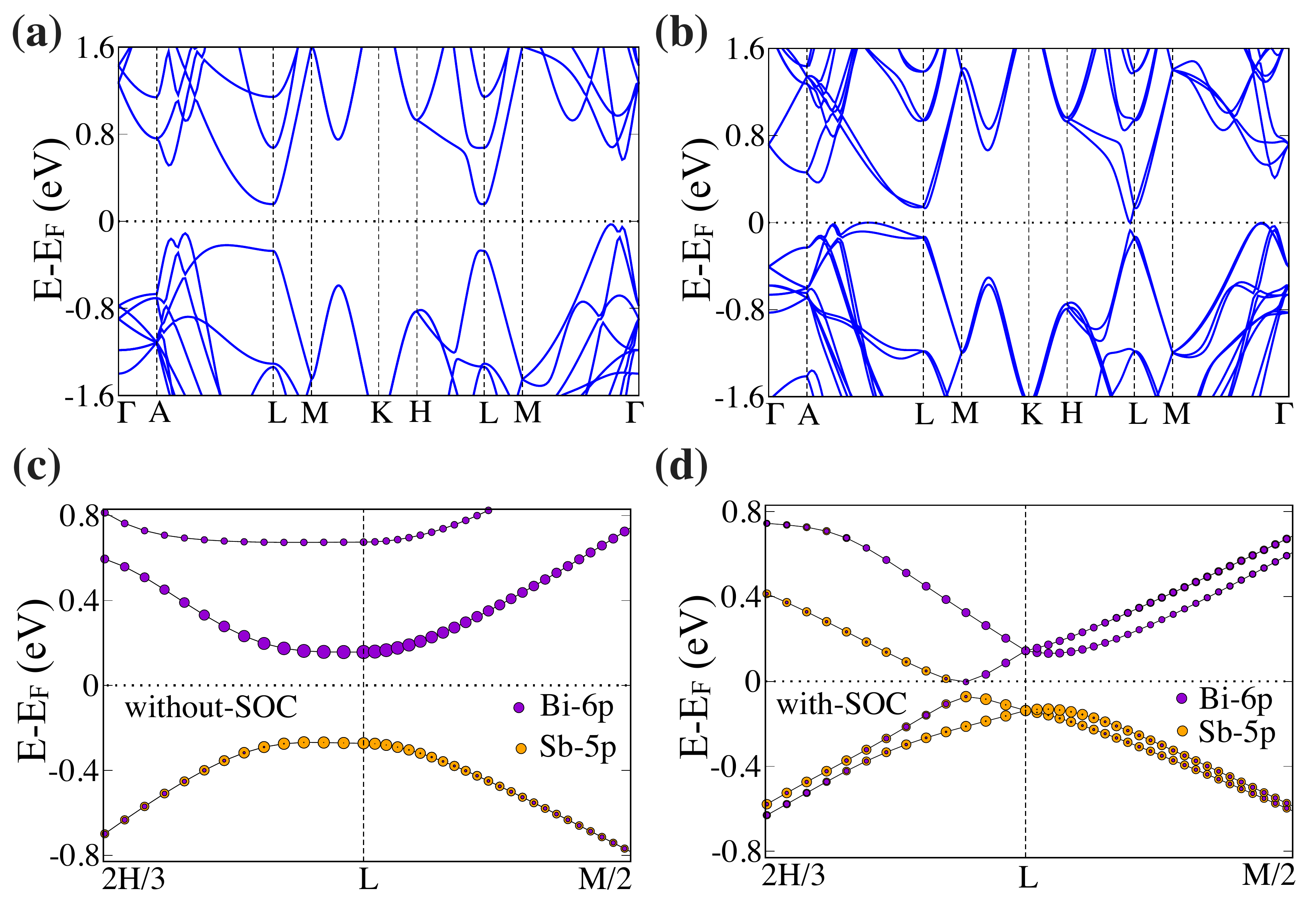}
 \caption{(Color online) Bandstructure (a) without-SOC and (b) with SOC. The effect of large SOC can be observed in the spin-splitting of bands near Fermi-level. Projection of Bi-$6p$ and Sb-$5p$ orbitals on bands close to the L-point is shown in (c) without-SOC and (d) with-SOC. Violet circles correspond to the projection of Bi-$6p$ orbitals and orange circles represent the projection of Sb-$5p$ orbitals. Band-inversion can be clearly observed along H $\rightarrow$ L path.\label{fig:1}}
 \end{figure}

In the \emph{R3m} space group, BiSb has a layered crystal structure with alternative layers of Bi and Sb atoms oriented along the $c$-axis of the rhombohedral unit-cell (see crystal structure in fig. 4a). Two alternative layers of Bi and Sb form a bilayer, which is repeated along the $c$-crystallographic direction of the unit-cell. Lattice parameters of the fully relaxed unit-cell are; $a$ = $b$ = 4.51  {\AA}, $c = 11.92 $  {\AA} and the angles are; $\alpha$ = $\beta$ = 90$^o$, $\gamma$ = 120$^o$. Additional details regarding crystal structure are given in the supplemental information. To study the dynamical stability of BiSb, we performed phonon calculations using Density Functional Perturbation Theory (DFPT) as implemented in ABINIT code \cite{gonze2002first, gonze2005brief, gonze2009abinit}. We used different approximations for exchange-correlation functional (LDA, LDA+SOC, PBE and PBE+SOC) to compute phonon modes. We obtained positive phonon frequencies for all four cases, which indicates that $R3m$ phase of BiSb is vibrationally stable at the level of DFPT calculations (supplemental information). However, we see a significant change in the phonon spectra due to the presence of SOC, which is in agreement with similar work reported for pristine Bi and Sb \cite{BiSOC_PRB2007, BiSOC_PRL2007, SbSOC_PRB2008}. Furthermore, we found that phonon frequencies are positive until 10.0 GPa external applied pressure. Beyond 10.0 GPa pressure phonon modes start becoming softer, which is indicative of a structural phase transition at higher pressures. To further ensure the stability of the crystal structure, we have performed calculations to obtain elastic constants \cite{Hill1952, Pugh1954}. Our calculations yield that BiSb system satisfies the mechanical stability criteria \cite{BornHuang1954, Shi2009} for a rhombohedral cell (details can be found in the supplemental information). Therefore, phonon and elastic constants calculations confirm the vibrational and mechanical stability of the BiSb system. We also find that this structure is robust against the structural and atomic disorders. 

Bulk bandstructure is calculated with and without-SOC. Figure 1(a) and 1(b) represent the calculated electronic bandstructure along the high symmetry lines of the hexagonal Brillouin zone. From bands without-SOC (fig. 1a), one can see that BiSb is an indirect bandgap semiconductor with indirect energy gap of 0.16 eV. The direct bandgap is located at the L-point (0.5, 0.0, 0.5) of the Brillouin zone with an energy gap of 0.43 eV. In the presence of SOC, each band of BiSb shows spin-splitting (fig. 1b) and the band-inversion occurs at the L-point. It is worthwhile to note that the spin-degeneracy of bands still exits at the L-point (Kramers' point), which confirms that BiSb breaks the I-symmetry but preserves the TR-symmetry. For TR-symmetric systems, lack of I-symmetry is an essential requirement for realization of WSM phase. Thus, this system fulfills the fundamental requirement to be in the WSM phase. The energy gap between the conduction band bottom (CBB) and valence band top (VBT) decreases due to the presence of SOC, particularly close to the L-point it has more pronounced changes. However, there still exists a small direct energy bandgap at the L-point along with indirect energy gap along the A-L path. 

Fig. 1c (1d) shows the $p$-orbital projections of Bi and Sb atoms on the bands close to the L-point for without-SOC (with-SOC) case. Violet circles represent Bi-$6p$ orbitals and orange circles represent Sb-$5p$ orbital projections. The diameter of circles is proportional to the band weights. For bands without-SOC, the states near the L-point of CBB are mainly composed of Bi-$6p$ orbitals. On the other hand, only Sb-$5p$ orbitals show their signatures close to the L-point of the VBT (Fig. 1c). When we take SOC into account, the SOC Hamiltonian causes a further decrease in the energy of the CBB state and increase in the energy of VBT state. Consequently, CBB state is pushed down and VBT state is pushed up, which eventually results in the inversion of the bands at the L-point. The strong SOC of Bi and Sb atoms is responsible for the bands-inversion. The band-inversion process is more evident along H $\rightarrow$ L direction. Such a band-inversion process yields topologically protected conducting surface states in the bulk insulating system. Thus, the bandstructure calculations reveal the non-trivial topological nature of bands in the BiSb system.

In 2014, R. Okugawa and S. Murakami proposed an effective model Hamiltonian to explain the evolution of Fermi-arcs and the WSM phase transition in I-asymmetric systems \cite{Okugawa2014}. In their model, they introduce a control parameter $m$ in the effective Hamiltonian to tune the bulk bandgap. By slowly varying $m$, one can close the bulk bandgap and realize gapless Weyl points at discrete points in the Brillouin zone. For I-asymmetric systems, these gapless Weyl points exist for a finite range of the control parameter $m$. On the other hand, for I-symmetric systems, Weyl points appear at a particular value of $m$ (say $m_o$) via pair-creation of monopoles and antimonopoles. With further increase in $m$ $(m > m_o)$, pair-annihilation occurs and the Weyl points disappear. Thus, for I-symmetric systems, the WSM phase lies exactly at the intermediate boundary of two distinct phases of topological insulator or normal insulator \cite{Murakami2007}.

\begin{figure}[ht!]
 \centering
 \includegraphics[width=7.5cm, keepaspectratio=true]{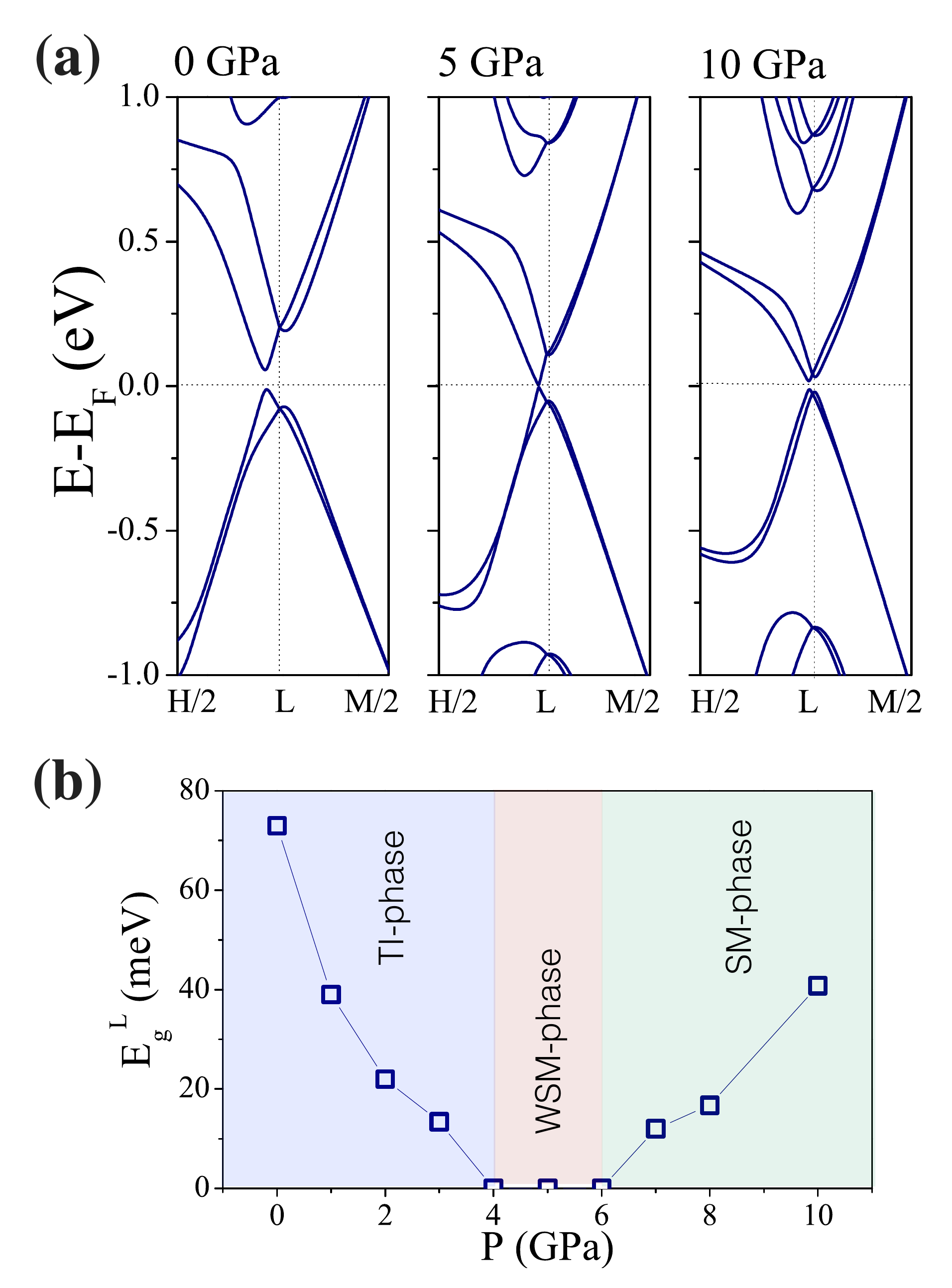}
 \caption{(Color online) Figure 2(a) represents the evolution of the electronic band structure (with-SOC) for different values of applied isotropic pressure, P = 0.0, 5.0, and 10.0 GPa. We start from a crystal system with clear electron bandgap, and as we increase the pressure, the direct electron bandgap along the H/2 $\rightarrow$ L path starts decreasing. The CBB and VBT touch each other at a pressure around 4.0 GPa, with the appearance of gapless Weyl nodes in k-space. The electron bandgap remains zero between 4.0 -- 6.0 GPa pressure range, and the system maintains the gapless Weyl nodes in this pressure range.  Beyond 6.0 GPa pressure the bandgap reopens again. Figure 2(b) depicts the change in the direct electron bandgap along H/2 $\rightarrow$ L direction as a function of applied pressure, P. The transitions of the electronic behavior of the material as a function of the isotropic pressure is observed, going from a topological insulator phase (TI-) to the Weyl SM phase (WSM-) and then, to the semimetal state (SM-phase). \label{fig:2}}
 \end{figure}

In the present work, we use isotropic pressure (P) as the control parameter to tune the bulk bandgap. Figure 2(a) shows the evolution in the direct bandgap along H $\rightarrow$ L direction for different values of applied pressure. The direct bandgap along H $ \rightarrow $ L direction ($E_g^L$) decreases as we increase pressure up to 3.0 GPa (fig. 2b). Interestingly, with further increase in pressure, $E_g^L$ becomes zero at $4.0$ GPa pressure and remains zero until 6.0 GPa. In this pressure range, the CBB and the VBT accidentally touch each other along H $\rightarrow$ L direction (see fig. 2(a) for 5.0 GPa case). The dispersion of the bands close to the band touching point is linear in $\textbf k$ and the band touching point is not located at any Kramers' points or along any high-symmetry direction of bulk Brillouin zone. These facts give us signatures of a Weyl semimetallic phase in 4.0 - 6.0 GPa pressure range. When we further increase the external pressure, we observe that the direct bandgap $E_g^L$ reopens. However, the system becomes semimetallic due to crossing of the Fermi-level by other topologically trivial bands. Thus, a pressure induced non-trivial insulator-WSM-semimetal phase transition takes place in the present BiSb system. A similar pressure induced phase transition has been reported for trigonal Te and Se systems \cite{HirayamPRL2015}. More details about the Weyl semimetallic phase occurring at 5.0 GPa pressure are given below. 

\begin{figure*}[htb!]
 \centering
 \includegraphics[width=17.5cm, keepaspectratio=true]{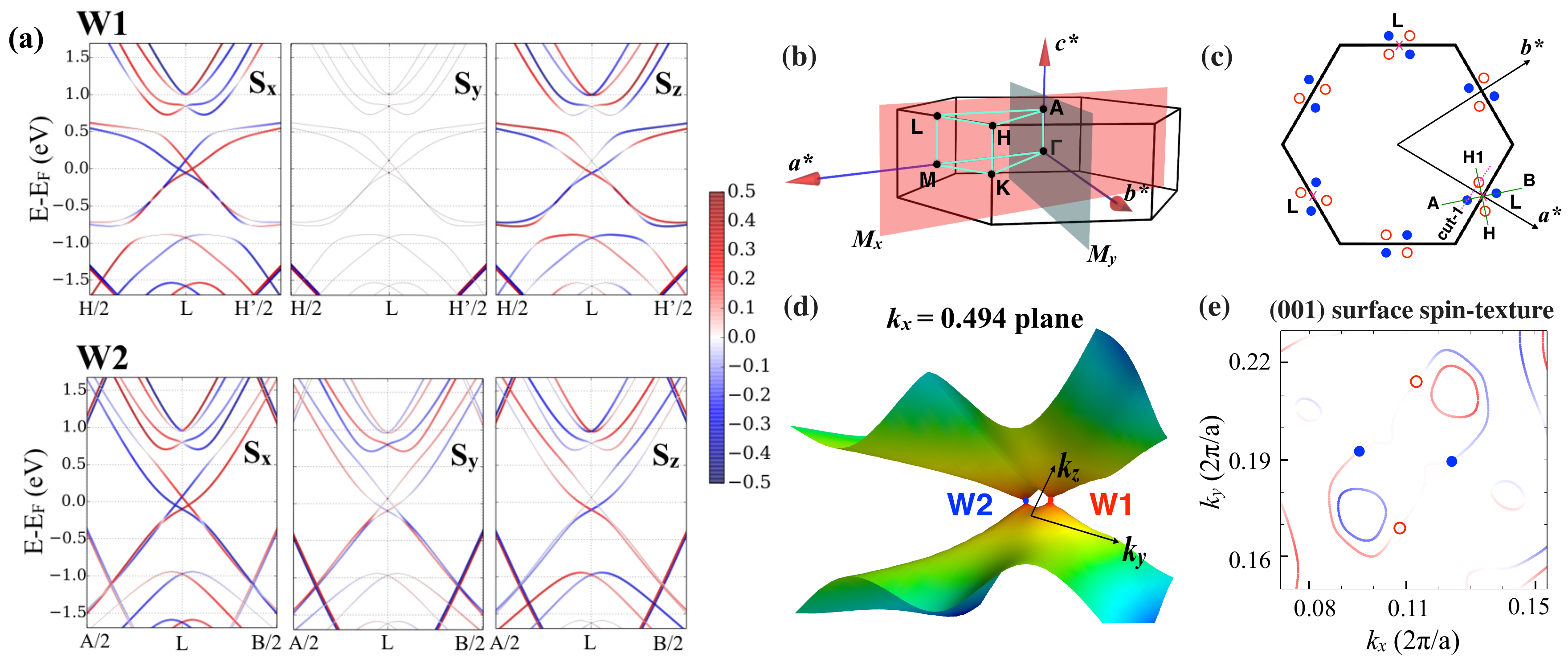}
 \caption{(Color online) Figure 3(a) shows the spin-projections on bands along H/2 $\rightarrow$ L $\rightarrow$ H'/2 (top) and A/2 $\rightarrow$ L $\rightarrow$ B/2 (bottom) directions for 5.0 GPa applied pressure (see figure 3(c) for directions in k-space). Red and blue color represents projections of up and down spin-orientations, respectively (see the color scale). Thin black lines show the original bands. We notice that bands near L-point are spin non-degenerate and VBT and CBB touch each other twice in the vicinity of the L-point. These spin non-degenerate band touching points represent Weyl points. Weyl points along H/2 $\rightarrow$ L $\rightarrow$ H'/2 direction (top) are labelled as W1 while Weyl points along A/2 $\rightarrow$ L $\rightarrow$ B/2 direction (bottom) are labelled as W2. The k-space coordinates of A and B points are ($\frac{1}{3}$, -$\frac{1}{3}$, $\frac{1}{2}$) and ($\frac{2}{3}$, $\frac{1}{3}$, $\frac{1}{2}$), respectively. It is important to notice that W1 Weyl points are located near the Fermi-level, however, W2 Weyl points are located slightly below the Fermi-level. Figure 3(c) shows locations of all Weyl points situated at $k_z$ = $\frac{\pi}{c}$ plane of the hexagonal Brillouin zone shown in figure 3(b). Solid blue and hollow red circles represent Weyl points with opposite chirality. Figure 3(d) shows the dispersion of bands in $k_x$ = 0.494 plane ($k_{y}-k_{z}$ plane passing through cut-1 in fig. 3c). Two gapless W1 and W2 Weyl points can be clearly seen in this figure. Figure 3(e) shows the spin-texture for (001) surface of BiSb. The reminiscence of Fermi-arcs connecting two Weyl points can be noticed. \label{fig:3}}
  \end{figure*}

Figure 3(a) shows the projection of spin-components ($S_{x}$, $S_{y}$, and $S_{z}$) on bands for 5.0 GPa applied pressure. We observe similar behavior of spin-projections on the bands for other values of P. Here, we have chosen the (001) direction as the quantization axis for all spin dependent calculations. Red (Blue) color represents the up (down) spin state. In figure 3(a), we calculate the band structure along two different directions in k-space:  H/2 $\rightarrow$ L $\rightarrow$ H'/2 and A/2 $\rightarrow$ L $\rightarrow$ B/2 (see figure 3(c) for directions). Along the H/2 $\rightarrow$ L $\rightarrow$ H'/2 direction, one can clearly notice that two non-degenerate gapless points are present in the vicinity of the L-point. Such gapless points in k-space are called Weyl points or Weyl nodes and unlike Dirac points they are 2-fold degenerate rather than 4-fold. Weyl points (W1) along H/2 $\rightarrow$ L $\rightarrow$ H'/2 direction are  located near the Fermi-level and are related by TR-symmetry (L-point as Kramers' point). Hence, these two points have the same chirality and topological charge. Bandstructure along A/2 $\rightarrow$ L $\rightarrow$ B/2 direction reveals presence of two more spin non-degenerate gapless points located slightly below Fermi-level. These two gapless points represent another set of Weyl points depicted by W2. W2 Weyl points are related by TR-symmetry and hence, they share the same chirality and topological charge. To determine the chirality of W1 and W2 Weyl points, we have calculated the Berry flux through a closed surface enclosing only one Weyl point at a time \cite{DavidBerryPhase1993, PythTB}. Our calculations reveal that W1 and W2 have opposite chirality and therefore, they carry opposite topological charges. Consequently, W1 behaves like a monopole and W2 behaves like an antimonopole in k-space. These two opposite Weyl points are located at different energy values. Existence of such opposite Weyl points, located at different energies, is attributed to the broken I-symmetry of the present BiSb system. Since W1 and W2 are separated in momentum space as well as in energy, it is impossible to realize a nodal semimetallic phase by tuning the Fermi-level and there always exists a state with separate electron and hole Fermi-surfaces. Weyl points exist in the system as long as the electron Fermi-surface is separated from the hole Fermi-surface. Such a property of WSM is very interesting as it could yield chiral magnetic effect and novel topological transport properties in the system. In recent years, these properties have attracted much theoretical interest \cite{Fukushima2008, Grushin2012, Aji2012, ZyuzinPRB2012, Zyuzin2012, Hosur2013, Vazifeh2013, Chang2015}. 

Using the symmetry of the crystal the position and chirality of all other Weyl points can be determined. Figure 3(c) shows the location of all monopoles and antimonopoles present at the $k_z = \frac{\pi}{c}$ surface of the hexagonal Brillouin zone. In reciprocal space, coordinates of the W1 and W2 points are: (0.494$\frac{2\pi}{a}$, -0.012$\frac{2\pi}{b}$, 0.500$\frac{2\pi}{c}$) and (0.494$\frac{2\pi}{a}$, 0.012$\frac{2\pi}{b}$, 0.500$\frac{2\pi}{c}$), respectively. To further confirm the existence of two Weyl points in the k$_x$ = 0.494 plane, we have calculated the band dispersion in k$_y$-k$_z$ plane (at constant k$_x$ = 0.494) near W1 and W2 points (see plane along cut-1 in figure 3c). Two gapless points can be clearly noticed in figure 3(d). One point is located at positive k$_y$ and another point is located at negative k$_y$, but both are located at the same k$_z$. There are three equivalent L-points in the hexagonal Brillouin zone and two Weyl points (W1 and W2) are located near each L-point. We further exploit $M_{x}$ and $M_{y}$ mirror symmetry operations to find the pairs of Weyl points located near each L-point.  Thus, in total there are 12 Weyl points, 6 monopoles and 6 antimonopoles, in the first Brillouin zone (figure 3c).

Observation of Fermi-arcs connecting two Weyl points is crucial to confirm the prediction of a WSM phase. For this purpose, we have performed the spin-texture calculations at (001) surface of a BiSb-slab \cite{PyProcar}. The slab was constructed from the primitive cell of BiSb. We observe two trivial Fermi-circles along with two open Fermi-curves [Fig. 3(e)]. These open Fermi-curves are reminiscent of the Fermi-arcs connecting opposite Weyl points \cite{Belopolski2016}. 

Having established the existence of WSM phase in BiSb, we will finally comment on the possibility to control the chirality of Weyl points. It is important to note that the experimental control of the spin-related properties stands as one of the key features in future devices and applications. The coupling between electric polarization and SOC in novel class of FERSC materials promises to solve this issue. Furthermore, the recent experimental discovery of ferroelectric-metals has generated huge interest in polar materials \cite{Shi2013, Nicole2016}. The polar ferroelectric-metals simultaneously exhibit two self-contradicting properties: polarity and metallicity. This particular behavior has been theoretically explained as follows: in ferroelectric-metals the free electrons screen out the long-range electrostatic forces favoring a polar structure with a net dipole moment \cite{Nicole2016}. The interplay between SOC and polarity could yield exotic quantum phenomenon in polar ferroelectric-metals \cite{bauer2012non, DaniloPRL2015}. Our calculations show that BiSb is a narrow-bandgap ferroelectric semiconductor at zero pressure. Also Bi and Sb atoms are known to exhibit strong SOC, as previously reported \cite{BiSOC_PRB2007, BiSOC_PRL2007, SbSOC_PRB2008}. Here, we take advantage of the large SOC and ferroelectric polarization to control the spin-related properties in the present BiSb system. 

At zero pressure, the BiSb material is a narrow-bandgap semiconductor which exhibits Rashba-like spin-splitting as observed in figure 2. In the $R3m$ phase, Bi and Sb atoms are displaced from the ideal rocksalt sites. The polar displacement of the Bi atoms along the direction of red arrows (see crystal structure shown in figure 4a) yields to a ferroelectric-like polarization in the BiSb system. The calculated unit cell polarization components (denoted as $P^s$ and obtained by means of the Berry-phase approach \cite{DavidBerryPhase1993, Vanderbilt2000147})  are $P_x^s$, $P_y^s$ and $P_z^s$ with values of -0.95, 1.64, and -5.31 $\mu$C$\cdot$cm$^{-2}$, respectively. It is worth mentioning that we can switch the polarization by reversing the direction of polar displacement of Bi-atoms (along blue arrows in figure 4b) from the ideal rocksalt sites \cite{Sante_GeTe2013, Silvia2014}. We notice that the spin-polarization of all three spin components of Rashba-like bands at 0.0 GPa gets inverted when we switch the direction of the ferroelectric polarization. This, as expected from the ferroelectric and SOC coupling terms in the Rashba Hamiltonian \cite{rashba1984}, makes BiSb (in $R3m$ phase) a FERSC material at ambient conditions. After this process, we apply 4.0 GPa isotropic pressure to reach the WSM phase with inverted spin-polarized bands. Figure 4 shows dispersion of bands along W1 and W2 Weyl points for original and inverted polarization cases. The shape of bands and the location of Weyl points in energy and momentum space is exactly the same for both $P_\uparrow^s$ and $P_\downarrow^s$ polarizations. However, after inverting the polarization we do notice a change in the spin-projected bands. We found that the spin-projection on the bands changes in spin-orientation which is depicted by switching of the up and down $S_x$ spin-components (figure 4). We observe similar behavior for $S_y$ and $S_z$ band spin-projections. The band spin-switching happens along both k-directions containing W1 and W2 Weyl points when the polarization is reversed. As a result, the topological charge and chirality of Weyl points attain opposite values after inverting  polarization.  This points to the possibility of switching of the monopole and antimonopole charges in a WSM via an electric field. This property is very important for designing new generation Weyltronics and spintronics devices.  Here, it is important to remark that the WSM-phase will retain the polar-like displacements similar to the case of  ``ferroelectric-metals"  \cite{Nicole2016}. This analysis demonstrates the coupling between the polarization and the SOC-degree of freedom in the TI- and WSM-phases in this BiSb crystal. In addition to this work, a recent study has also predicted the existence of a WSM phase in ternary hyperferroelectrics \cite{DiSante2016} by means of chemical doping and alloying. 

\begin{figure}[ht!]
 \centering
 \includegraphics[width=8.7cm, keepaspectratio=true]{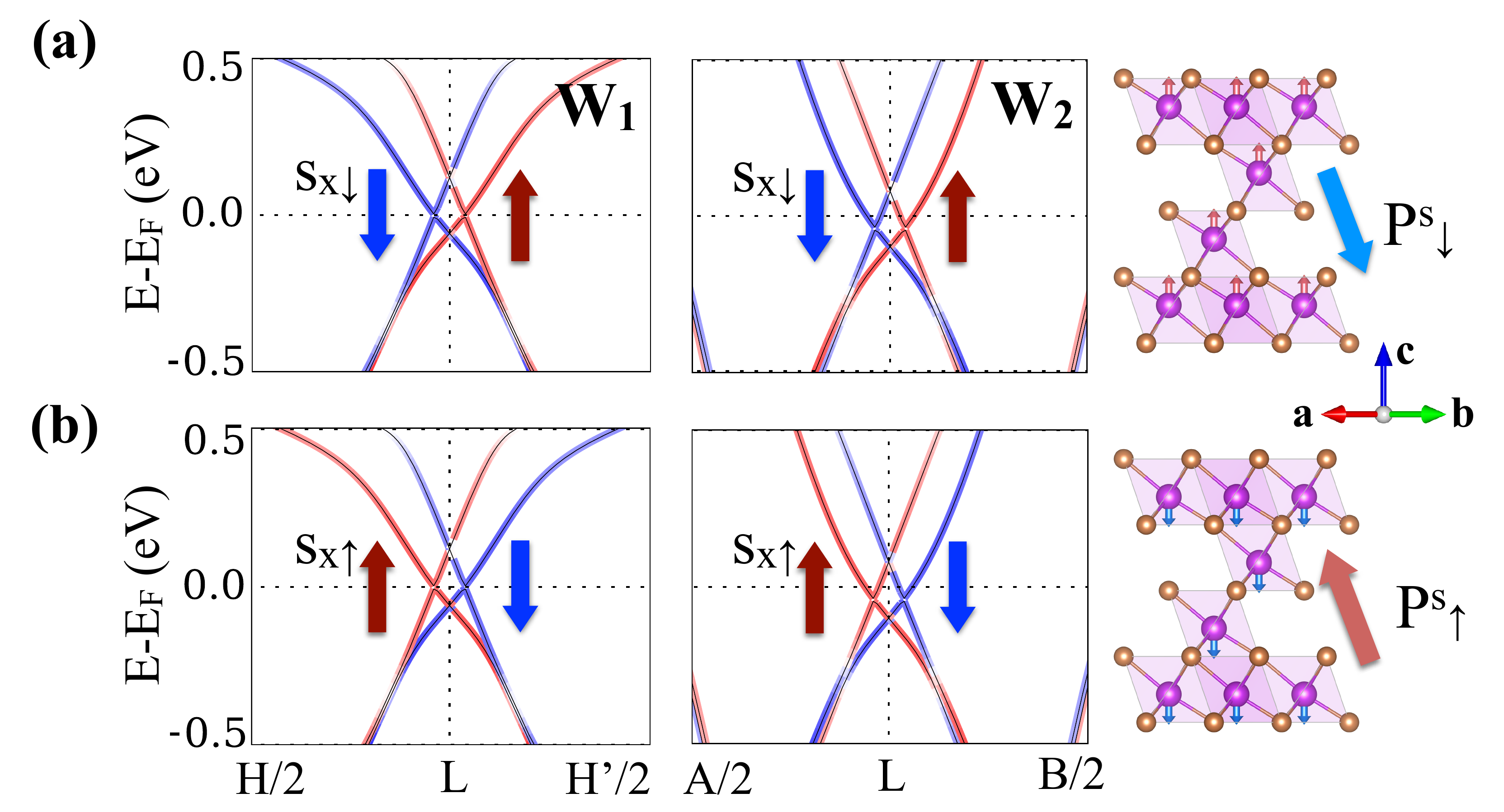}
 \caption{The spin-control and switching of Weyl points by means of polar-displacements (P$^s$) in the system can be observed. Top (Bottom) figure 4a (4b) shows the spin-projected band dispersions near W1 and W2 Weyl points for original (inverted) polarization cell. The change in the polarization is equivalent to the Bi-displacements along up or down directions as shown by the red and blue arrows. Bi and Sb atoms are shown in violet and orange colors, respectively.  \label{fig:4}}
 \end{figure}

In summary, we report a stoichiometric layered structure of BiSb which is enthalpically, vibrationally and mechanically stable at the DFT level. This structure has a bulk insulating gap at zero pressure with non-trivial band topology. Pressure dependent electron band structure calculations reveal the existence of a WSM phase in the pressure range 4.0 - 6.0 GPa. We have found 6-pairs of Weyl points at $k_z = \pi/c$ surface of the Brillouin zone. We clearly observe linear dispersion of spin non-degenerate bands near each Weyl point. Surface state calculations further confirm the presence of WSM phase in BiSb. Additionally, the WSM phase of this non-magnetic system relies on the broken inversion symmetry of the crystal rather than the time-reversal asymmetry. This feature makes this material very appealing and interesting to experimentalists. We further discussed the possibility to control the chirality and topological charge of the Weyl points by means of applied pressure and electric field. The existence of a polar-like Weyl semimetallic phase in this BiSb compound calls for further studies on the underlying exotic quantum phenomena \cite{Bauer2012}. \\

\begin{center}
\textbf{Computational Details}
\end{center}

We used the projected augmented wave (PAW) method implemented in the VASP code to perform all DFT-based first principle calculations \cite{Kresse1996, Kresse1999}. We considered fifteen valence electrons of Bi ($5d^{10} 6s^2 6p^3$ ) and five valence electrons of Sb ($5s^2 5p^3$) in the PAW pseudo-potential. For exchange-correlation potential, we used exchange-correlation functional parameterized by Perdew-Burke-Ernzerhof \cite{Perdew1996}. For all calculations presented in this paper, we used a conventional unit-cell of BiSb (space group 160: \emph{R3m}), containing 3 atoms of Bi and 3 atoms of Sb. This cell was obtained after a systemic study of the low energy phases of Bi$_{1-x}$Sb$_x$ (0 $<$ \emph{x} $<$ 1) compounds by using minima hopping structure search method \cite{Goedecker2004, Amsler2010}. An expanded set of results including other stable stoichiometries will be published elsewhere. We used 650 eV as kinetic energy cut-off for the plane wave basis set and a Gamma 10$\times$10$\times$10 \emph{k}-point mesh has been employed for all DFT based first principle calculations. For convergence of electronic self-consistent calculations, a total energy difference criterion was defined as 10$^{-8}$ eV. For ionic relaxations, the cell was relaxed, first in the internal coordinates and then in the volume until the total energy difference between two consecutive steps was smaller than 10$^{-8}$ eV and the total residual forces were less than 10$^{-5}$ eV/{\AA}. Spin-orbit interaction was included for all ionic relaxations. We use ABINIT code to calculate phonon spectra \cite{gonze2002first, gonze2005brief, gonze2009abinit}. We have used a q-mesh of size $4\times4\times4$ for all phonon calculations.  Band-structure calculations were done (i) with spin-orbit coupling (SOC) and (ii) without SOC. For all pressure dependent electronic structure calculations, we employ isotropic pressure ranging from 1 GPa to 10 GPa on the original unit cell. To visualize the open Fermi-arcs, we have calculated the spin-texture at (001) surface of a BiSb-slab \cite{PyProcar}. This 55.6 {\AA} thick slab consists 14-bilayers of BiSb and a vacuum layer of 12.0 {\AA} thickness. VESTA \cite{vesta2008}, PyProcar \cite{PyProcar} and MAYAVI \cite{mayavi2011} softwares were used to make some figures. 

\begin{center}
\textbf{Acknowledgement}
\end{center}
This work used the Extreme Science and Engineering Discovery Environment (XSEDE), which is supported by National Science Foundation grant number OCI-1053575. Additionally, the authors acknowledge the support from Texas Advances Computer Center (TACC) and Super Computing System (Mountaineer and Spruce) at WVU, which are funded in part by the National Science Foundation EPSCoR  Research Infrastructure Improvement Cooperative Agreement 1003907, the state of West Virginia (WVEPSCoR via the Higher Education Policy Commission) and WVU. A. H. Romero and S. Singh also acknowledge the Donors of the American Chemical Society, Petroleum Research Fund for partial support of this research under contract 54075-ND10 and NSF with the DMREF-NSF project 1434897. F. Mu\~noz acknowledges support from Fondecyt under grant 1150806 and Center for the development of Nanoscience and Nanotechnology CEDENNA FB0807. S. Singh thanks Dr. Guillermo Avenda\~no-Franco for assisting in preparation of some figures. 

\bibliographystyle{ieeetr}

\end{document}